\PassOptionsToPackage{sort&compress}{natbib}
\documentclass[preprint,12pt]{elsarticle}
\usepackage{amssymb}
\usepackage{amsmath}

\usepackage{graphicx,xcolor}

\journal{Physics of the Dark Universe}

\begin{document}

\title{Collective Excitations of 
Self-Gravitating Ultralight Dark Matter Cores}

\author{Luca Salasnich$^{1,2}$ and Alexander Yakimenko$^{1,3}$}

\address{$^{1}$Dipartimento di Fisica e Astronomia ``Galileo Galilei'',
Universit\`a di Padova, and INFN, Sezione di Padova,
Via Marzolo 8, 35131 Padova, Italy \\
$^{2}$Istituto Nazionale di Ottica (INO) del Consiglio Nazionale
delle Ricerche (CNR), Via Nello Carrara 1, 50019 Sesto Fiorentino, Italy \\
$^{3}$Department of Physics, Taras Shevchenko National University
of Kyiv, 64/13, Volodymyrska Street, Kyiv 01601, Ukraine}

\begin{abstract}
A distinctive feature of ultralight bosonic dark matter is its ability to form a Bose–Einstein condensate with a dense, stationary configuration at the center of galactic halos. In this work, we investigate the internal dynamics of such cores by numerically solving the Bogoliubov–de Gennes equations within a fully self-consistent gravitational framework, accounting for both gravitational potential perturbations and local self-interactions. We demonstrate that the solitonic core supports a discrete spectrum of well-defined collective modes. These oscillations characterize the linear response of the core to perturbations and may influence various dynamical processes. We also discuss potential astrophysical implications of these excitations on galactic scales.
\end{abstract}

\maketitle
\section{Introduction}
The internal structure of galactic halos provides a powerful probe of the fundamental nature of dark matter, which remains one of the central open questions in modern physics. Among the possible candidates for the dark matter particle, ultralight bosons with cosmic-scale Compton wavelengths have attracted growing attention due to their ability to form coherent, self-gravitating configurations on kiloparsec  scales~\cite{chavanis_book,ferreira2021ultra,baldeschi1983massive,jackson2023search,PhysRevD.103.123551}. 

A key prediction of such models is the emergence of a Bose–Einstein condensate (BEC) on galactic scales, with distinctive dynamical behavior in the central regions of ultralight dark matter (ULDM) halos.
These BEC cores exhibit a smooth transition to a more diffuse outer envelope \cite{mocz2017galaxy}. The inner region is often modeled as a solitonic core, while the outskirts resemble a standard cold dark matter halo. Although this core-halo structure is a robust feature, several aspects of the outer halo including its detailed density profile and mass distribution, remain under active investigation~\cite{de2022accurate,korshynska2023dynamical,harko2012finite,PhysRevD.100.083022}. \textcolor{black}{Early theoretical insights into the gravitational instability of scalar fields, including axion-like particles, were presented in Ref.~\cite{Khlopov1985}, which anticipated the formation of localized self-gravitating condensates.}

While a pure BEC soliton is a stationary solution of the Gross–Pitaevskii\-Poisson (GPP) system, realistic halos exhibit significant time-dependent dynamics. In particular, Ref. \cite{PhysRevD.98.043509} showed that the central density of an isolated ULDM halo can oscillate substantially, deviating from the ideal soliton profile. Such oscillations, potentially observed in ultrafaint dwarf galaxies, may contribute to star cluster heating and long-term expansion \cite{PhysRevLett.123.051103,PhysRevD.103.103019}. In addition to these internal modes, solitonic cores can undergo a confined random walk driven by wave interference, producing fluctuations in the external gravitational field that affect the dynamics of satellite galaxies~\cite{PhysRevLett.124.201301}. These phenomena have motivated numerous theoretical studies on the stability, structure, and excitations of perturbed ULDM cores~\cite{PhysRevD.103.023508,PhysRevD.105.023512,PhysRevD.105.103506,PhysRevD.109.043516,PhysRevD.97.103523,bernstein1998eigenstates,tod1999analytical,chowdhury2021random,PhysRevD.109.103518,PhysRevD.110.023504,asakawa2023collective,liu2023coherent,PhysRevD.104.083532,harrison2002numerical,schroven2018self,PhysRevD.84.043531,LTP2021,PhysRevD.108.023503,PhysRevD.111.023006,sakstein2024dark,PhysRevD.101.081302}, with growing interest in identifying their potentially observable consequences.

In particular, Refs. \cite{PhysRevD.103.023508,PhysRevD.105.023512,PhysRevD.105.103506} examine the dynamics of the Schr\"odinger-Poisson (SP) system in the absence of local self-interaction. This framework describes Fuzzy Dark Matter (FDM), a model of ultralight bosonic particles that gives rise to quantum states on galactic scales, exhibiting wave-like behavior. These studies numerically determined the eigenstates of the SP system using an averaged gravitational potential, {assumed to be constant in time}. Work \cite{PhysRevD.110.023504} studied perturbations in a hybrid model which account both a coherent condensate state and an incoherent particlelike state. It was found in Ref.~\cite{PhysRevD.109.043516} that solitons, driven by repulsive self-interactions within extended halos of scalar-field dark matter, form rapidly and can constitute a significant fraction of the total mass. In addition to analyzing eigenmodes in a stationary potential, that study also investigated how different modes perturb the gravitational potential.
Ref.~\cite{asakawa2023collective} offers valuable insight into the nonlinear dynamics of self-gravitating BECs, demonstrating breathing and quadrupole mode excitation and the emergence of anisotropic deformation via full 3D time-dependent simulations of the GPP equations. Recent work \cite{PhysRevD.109.103518} employed a perturbative method to examine the {radial} oscillatory dynamics of solitonic cores in fuzzy dark matter, revealing universal features across interaction regimes. While these studies have advanced our understanding of ULDM core dynamics, they typically neglect the self-consistent dynamical coupling between density and gravitational potential perturbations, which are comparable in magnitude and crucial for describing core evolution.

Addressing these limitations requires solving the full Bogoliubov-de Gen\-nes (BdG) problem, which systematically captures the stability and collective excitations of the solitonic state via linear perturbations around equilibrium. However, the integro-differential structure of the BdG equations poses significant technical challenges due to the substantially nonlocal nature of the gravitational response.
In this work, we derive and numerically solve the complete BdG equations for a self-gravitating BEC soliton, fully accounting for perturbations in both the density and gravitational potential. Our analysis reveals how gravity and self-interactions shape the internal oscillation modes of the core, providing a deeper understanding of its dynamical behavior and offering potential links to observable features in galactic systems.

The paper is organized as follows. Section II presents the basic equations that govern the dynamics of self-gravitating BECs. In Section III, we recall the general properties of stationary solitonic cores. In Section IV, we introduce the BdG equations and describe the numerical methods used to solve them. Section V focuses on the breathing modes of the solitonic core. 
Section VI examines the astrophysical implications of collective modes, with emphasis on their dynamical effects and potential observability.
Finally, Section VII summarizes our findings and outlines potential directions for future research.

\section{Model}
On galactic scales, ultralight bosonic dark matter is effectively described by a nonrelativistic complex wavefunction $\tilde{\Psi}(\tilde{\mathbf{r}},\tilde{t})$ coupled to a Newtonian gravitational potential $\tilde{\Phi}(\tilde{\mathbf{r}},\tilde{t})$. In the weak-field regime and on scales well below the cosmological horizon, the Einstein–Klein–Gordon equations reduce to the Gross–Pitaevskii–Poisson (GPP) system, which accurately captures the self-gravitating dynamics of a coherent bosonic medium relevant to halo structure formation~\cite{PhysRevLett.85.1158, chavanis_book, ferreira2021ultra}:
\begin{align}
i\hbar \frac{\partial \tilde{\Psi}}{\partial \tilde{t}} &= \left( -\frac{\hbar^2}{2m} \nabla^2 + m\tilde{\Phi} + g|\tilde{\Psi}|^2 \right) \tilde{\Psi}, \label{eq:GPE_dim} \\
\nabla^2 \tilde{\Phi} &= 4\pi G m |\tilde{\Psi}|^2. \label{eq:Poisson_dim}
\end{align}
The wavefunction is normalized as
\begin{equation}
\int|\tilde{\Psi}|^2\, d^3 \tilde{\mathbf{r}} = N, \label{eq:Ndef}
\end{equation}
where $N$ is the total number of bosons in the solitonic core, $m$ is the particle mass, and $g = 4\pi\hbar^2 a_s / m$ quantifies local self-interactions via the $s$-wave scattering length $a_s$. We focus on the repulsive case ($a_s > 0$), which stabilizes the core against collapse and governs its structure in the Thomas–Fermi regime~\cite{chavanis_book}. In our model, dark matter is non-relativistic and extremely cold, such that the contribution of a thermal, non-degenerate component is negligible in the core region. Under these conditions, Eqs.~(\ref{eq:GPE_dim}), (\ref{eq:Poisson_dim}) accurately describe a gravitationally bound Bose–Einstein condensate.

It is instructive to reformulate this model in equivalent hydrodynamic form (see, e.g. \cite{chavanis_book}), applying the Madelung transformation, \(\tilde{\Psi} = \sqrt{\tilde{\rho}/m}\, e^{iS/\hbar}\), where \(\tilde{\rho} = m|\tilde{\Psi}|^2\) is the mass density and \(\tilde{\mathbf{v}} = \nabla S / m\) is the velocity field for
an irrotational flow. Substitution into Eqs.~\eqref{eq:GPE_dim}–\eqref{eq:Poisson_dim} yields the continuity equation,
\begin{equation}
\frac{\partial \tilde{\rho}}{\partial \tilde{t}} + \nabla \cdot (\tilde{\rho} \tilde{\mathbf{v}}) = 0, \label{eq:continuity}
\end{equation}
and the Euler-type equation,
\begin{equation}
\frac{\partial \tilde{\mathbf{v}}}{\partial \tilde{t}} + \left(\tilde{\mathbf{v}}\cdot\nabla\right)\tilde{\mathbf{v}} = -\nabla \tilde{\Phi} - \frac{1}{\tilde{\rho}} \nabla P -\frac{1}{m} \nabla Q, \label{eq:hydro_euler}
\end{equation}
where gravitational potential $\tilde{\Phi}$ satisfies Poisson equation (\ref{eq:Poisson_dim}), effective pressure $P=g\tilde{\rho}^2/(2m^2)$ corresponds to the local self-interaction and the last term is related to effective pressure with quantum potential
\begin{equation}
Q=-\frac{\hbar^2}{2m}\frac{\nabla^2\sqrt{\tilde{\rho}}}{\sqrt{\tilde{\rho}}}.
    \label{eq:Q}
\end{equation}
 These quantum effects, operating on astrophysical scales, arise from the underlying wave nature of the field and are well captured by the Schr\"odinger\-Poisson framework~\cite{chavanis_book,ferreira2021ultra}. 
 The fluid equation (\ref{eq:hydro_euler}) illustrates the interplay between attractive gravity, self-interaction (repulsive  for $a_s>0$), and repulsive wave dispersion (quantum pressure),  and allows the consideration of hydrostatic equilibrium ($\tilde{\textbf{v}}=0$). For example, in the noninteracting case ($a_s = 0$), the radial density profile $\tilde{\rho}(\tilde{r})$ of the stationary state was obtained numerically~\cite{chavanis_book,PhysRevD.84.043532,PhysRev.187.1767}. The radius of the configuration enclosing 99\% of the $N$ particles is given by $R_{99} \approx 9.95\, r_B / N$, where the gravitational analogue of the Bohr radius, $r_B$, is defined as
\begin{equation}
r_B = \frac{\hbar^2}{G m^3}.
\label{eq:r_B}
\end{equation}

We introduce dimensionless variables based on the characteristic spatial and temporal scales $L_*$ and $\Omega_*$, 
defined as
\begin{equation}
\Omega_* = \frac{\hbar}{m L_*^2},  \quad L_*^2=a_sr_B.
\end{equation}

Let dimensional quantities carry tildes, and define the rescaled variables as
\begin{equation}
\mathbf{r} = \frac{\tilde{\mathbf{r}}}{L_*}, \quad
t = \Omega_* \tilde{t}, \quad
\Psi = \frac{\tilde{\Psi}}{\psi_*}, \quad
\Phi = \frac{\tilde{\Phi}}{\varphi_*},
\end{equation}
with 
$\psi_*^2=N/L_*^3$ and $\varphi_*=Gm/a_s$.

In dimensionless form, the GPP system becomes
\begin{align}
i \frac{\partial \Psi}{\partial t} &= \left( -\frac{1}{2} \nabla^2 + \Phi + \gamma N_* |\Psi|^2 \right) \Psi, \label{eq:GPE_nondim} \\
\nabla^2 \Phi &= N_* |\Psi|^2, \label{eq:Poisson_nondim}
\end{align}
where $\gamma = +1$ for repulsive interactions and 
\begin{equation}\label{eq:N_star}
N_* = 4\pi N \sqrt{a_s/r_B}
\end{equation}
is the dimensionless particle number, which corresponds to the total mass of the core, \( M = mN = mN_* r_B / (4\pi L_*) \), in dimensional units. The wavefunction is normalized as
\begin{equation}
\int |\Psi(\mathbf{r},t)|^2\, d^3\mathbf{r} = 1.
\end{equation}


In what follows, all quantities are expressed in dimensionless units. This formulation enables a scalable analysis that is largely independent of the specific values of the bosonic particle parameters, making the results broadly applicable within the GPP framework. For instance, assuming an ultralight dark matter particle mass $m \sim 10^{-22} \,\mathrm{eV}$ and a characteristic scale $L_* \sim 1 \,\mathrm{kpc}$, we estimate an oscillation timescale of $\tilde{T}=2\pi/\Omega_*\sim 100\,\mathrm{Myr}$, consistent with cosmological simulations~\cite{PhysRevD.103.103019}. This supports the astrophysical relevance of the eigenmode timescales captured by our analysis.


Before analyzing the dynamical excitations, we first examine the stationary solitonic configurations that arise from the balance of gravitational attraction, quantum pressure, and repulsive self-interaction.

\section{Stationary solitonic core}
The system of equations (\ref{eq:GPE_nondim}) and (\ref{eq:Poisson_nondim}) allows the existence of stationary solitonic solutions with the wave function
\begin{equation}
\Psi(\mathbf{r},t) = \psi_0(r)e^{{-i\mu t}},   
\label{eq:stat_st}
\end{equation}
where $\mu$ is the chemical potential and $\psi_0(r)$ is the radial profile of the wave function.

\subsection{Numerical solitonic solutions}
The stationary equation for the dimensionless ground state wave function $\psi_0(r)$ of the bosons in the BEC state is as follows:
\begin{equation}
\hat{\mathcal{H}}_{l=0}\psi_0({r}) =0, 
\label{eq:StationaryGPE_dimless}
\end{equation}
where 
\begin{equation}
\label{eq:H_l_equation}
\hat{\mathcal{H}}_l=-\mu -\frac12\Delta_r^{(l)} + \gamma N_*\psi_0^2(r)+\Phi_0(r)    
\end{equation}
\begin{equation}
\Delta_r^{(l)}=\frac{d^2}{dr^2}+\frac{2}{r}\frac{d}{dr}-\frac{l(l+1)}{r^2},
\end{equation}
the gravitational potential for the spherically-symmetric solution is given as follows:
\begin{equation}\label{eq:Phi_0_integralform}
\Phi_0(r) = \Phi_0(0)+N_* \int_0^r\xi\psi_0^2(\xi)\left[1-\frac{\xi}{r}\right]d\xi,
\end{equation}
where
\begin{equation}
\Phi_0(0)=- N_*\int_0^{+\infty}{\xi}\psi_0^2(\xi)d\xi.    
\end{equation}

The normalization condition in dimensionless units is as follows:
\begin{equation}
\label{eq:normalized_dimless}
4\pi\int_0^{+\infty}\psi_0^2(\xi)\xi^2d\xi=1.
\end{equation}

We solve numerically the set of Eq. (\ref{eq:H_l_equation}), (\ref{eq:Phi_0_integralform}) of nonlinear integrodifferential equations
using the stabilized relaxation procedure similar to that employed in Refs. \cite{PRE2005,LTP2021}

\begin{figure}[ht]
    \centering
\includegraphics[width=\textwidth]{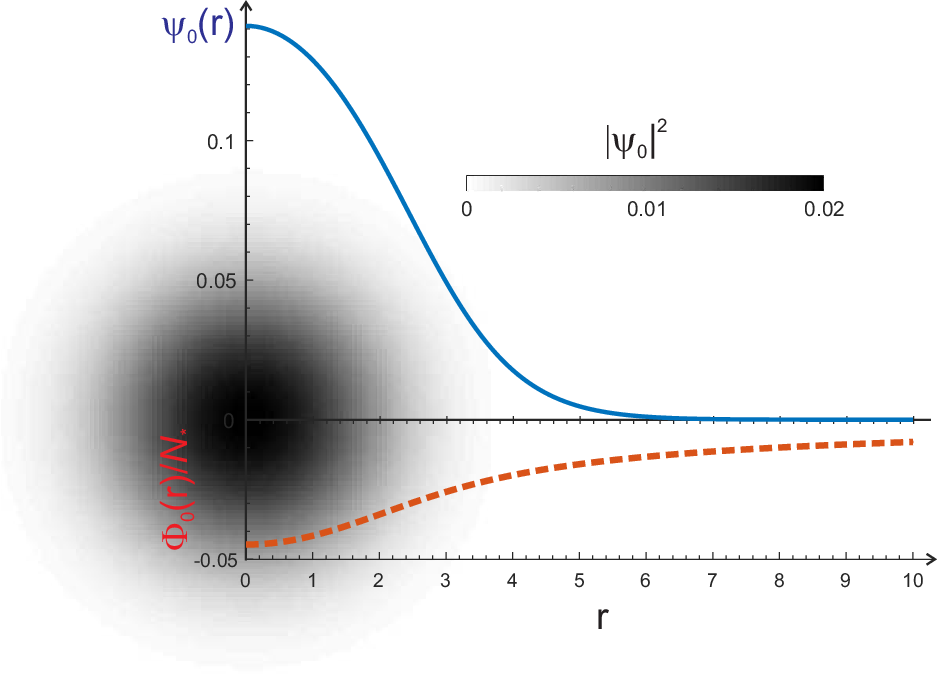}
    \caption{Numerical stationary ground-state solution in dimensionless units for $N_*=100$. The figure illustrates the radial profile of the condensate wave function $\psi_0(r)$ (solid blue line) and the normalized potential $\Phi_0(r)/N_*$ (dashed red line). The condensate density $|\psi_0|^2$ at $z=0$ plane is represented in the grey-scale density plot.}
\label{fig:Steady_statePsi_0Phi_0}
\end{figure}

The typical example of numerically obtained ground-state solution is presented in Fig. \ref{fig:Steady_statePsi_0Phi_0} for $N_*=100$, corresponding to a chemical potential $\mu = -2.2172$. The gray-scaled density plot shows the condensate density distribution, $|\psi_0|^2$, at $z=0$ plane, in dimensionless units. The radial profile of the wave function, $\psi_0(r)$, is depicted by the solid blue line, while the red dashed line represents the radial profile of the normalized gravitational potential, $\Phi_0(r)/N_*$, generated by the self-gravitating solitonic core. Note that at large distances, the gravitational potential exhibits Coulomb-like asymptotics, whereas in the central region, it is well approximated by a harmonic oscillator potential.

\begin{figure}[ht]
    \centering
    \includegraphics[width=0.75\textwidth]{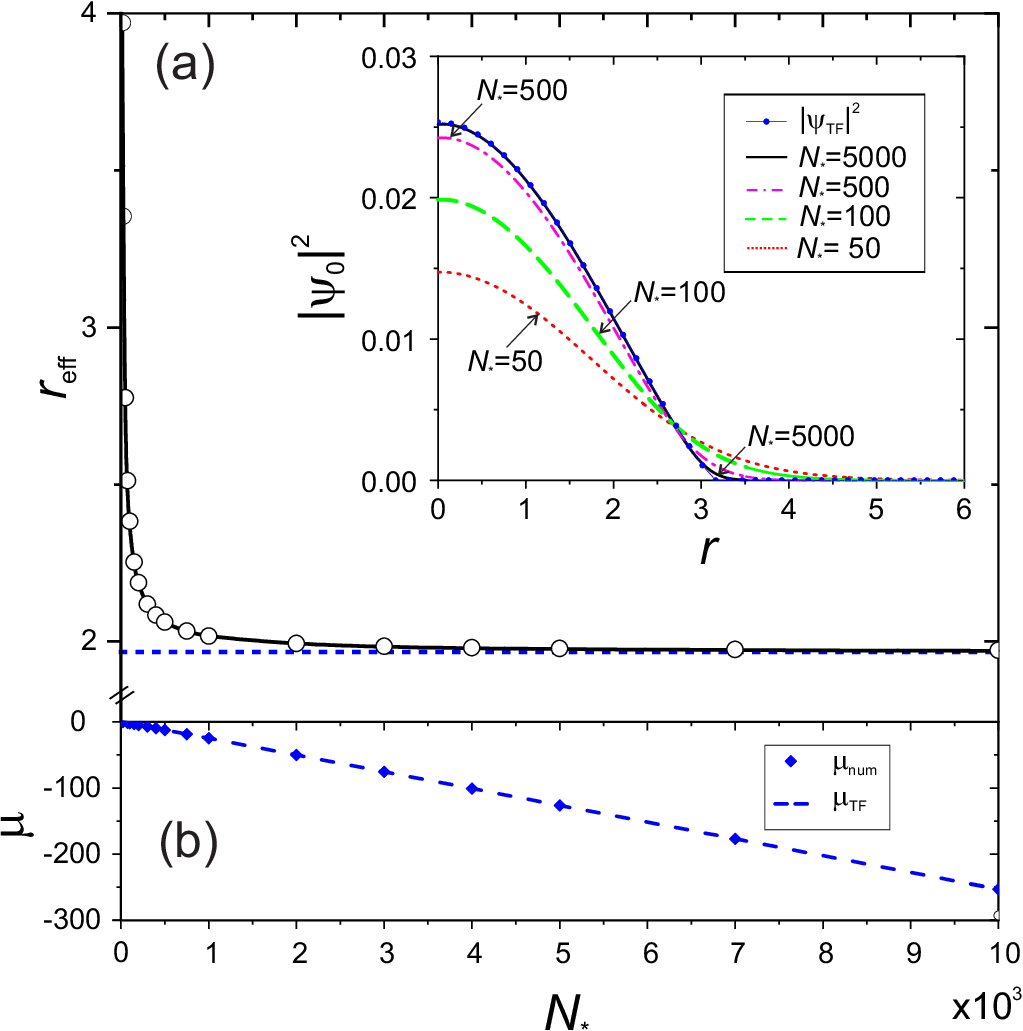}
    \caption{(a) Effective radius vs dimensionless number of bosonic particles $N_*$. The inset illustrates the radial profiles of the condensate density $|\psi_0|^2$ for different $N_*$. (b) chemical potential $\mu$ vs $N_*$. Shown are the results of numerical simulations (blue diamonds) and Thomas Fermi approximation (dashed blue line).}
    \label{fig:ProfilesPsiSq}
\end{figure}

Figure \ref{fig:ProfilesPsiSq} presents stationary solitonic solutions of Eq. (\ref{eq:StationaryGPE_dimless}) with the norm determined by Eq. (\ref{eq:normalized_dimless}) for various values of parameter $N_*$. Figure \ref{fig:ProfilesPsiSq} (a) shows the effective (averaged) radius, $r_{\textrm{eff}}$, defined by
\begin{equation} 
r^2_{\textrm{eff}} = 4\pi \int_0^{+\infty} \xi^4 \psi^2_0(\xi) d\xi. 
\end{equation} The inset in Fig. \ref{fig:ProfilesPsiSq} (a) depicts the corresponding radial profiles of the condensate density. Figure  \ref{fig:ProfilesPsiSq} (b) displays the chemical potential of the stationary state as a function of $N_*$. We further compare our numerically obtained stationary solitonic solutions with analytical estimates derived using the TF approximation.

\subsection{Thomas-Fermi approximation}

The influence of the local nonlinear term, corresponding to the repulsive self-interaction in the Gross-Pitaevskii equation, becomes increasingly significant with the growth of the dimensionless parameter $N_*$, as defined by Eq. (\ref{eq:N_star}). In Thomas-Fermi (TF) approximation neglecting the contribution of the quantum pressure in the equation of hydrostatic equilibrium \cite{chavanis_book,asakawa2023collective} it is easy to obtain the normalized condensate density:
\begin{equation}
    \label{eq:psi_TF}
    |\psi_{\textrm{TF}}(r)|^2=\frac{1}{4\pi^2}\left\{
\begin{array}{cc}
\frac{\sin r}{r},  & 0\le r \le \pi   \\
0 ,  & r >\pi 
    \end{array}
    \right.
\end{equation}
and corresponding gravitational potential:
\begin{equation}
    \label{eq:Phi_TF}
    \Phi_{\textrm{TF}}(r)=-\frac{N_*}{4\pi^2}\left\{
\begin{array}{cc}
1+\frac{\sin r}{r},  & 0\le r \le \pi   \\
\frac{\pi}{r},  & r >\pi.
    \end{array}
    \right.
\end{equation}
The TF radius of the solitonic core $R_{TF}=\pi$, effective radius of the solitonic core: $r_{\textrm{eff}}=\sqrt{\pi^2-6}\approx 1.967$, $\psi_{\textrm{TF}}(0)=\frac{1}{2\pi}$, $\Phi_{\textrm{TF}}(0)=-\frac{N_*}{2\pi^2}$. The chemical potential in TF approximation is: $\mu_{\textrm{TF}}=-\frac{N_*}{4\pi^2}$.

The inset in Fig.~\ref{fig:ProfilesPsiSq}(a) and Fig.~\ref{fig:ProfilesPsiSq}(b) compare the density distribution $|\psi_{\textrm{TF}}|^2$ and chemical potential $\mu_{\textrm{TF}}$ as functions of $N_*$ with corresponding numerical results. As $N_*$ increases, the radial profiles obtained from the numerical solution of the stationary GPP equations exhibit smooth convergence toward the TF limit. In particular, the effective core radius approaches the TF value $r_{\textrm{eff}}$ [indicated by the dashed blue line in Fig.~\ref{fig:ProfilesPsiSq}(a)] in the limit $N_* \gg 1$. For example, a solitonic core composed of bosons with mass $m = 10^{-22} \,\mathrm{eV}$ is well described by the TF approximation for $N_* = 500$, corresponding to a physical mass $M \approx 10^{10} M_\odot$ and TF radius $\tilde{R}_{\rm TF} \approx 1 \,\mathrm{kpc}$.

\section{Bogoliubov-de Gennes analysis of core excitations}
The important information on dynamics of the solitons
can be obtained from the analysis of small perturbations
of the stationary states. The basic idea of such a linear stability
analysis is to represent a linear perturbation as a superposition
of the modes with different angular symmetries.
\subsection{BdG equations for self-gravitating ULDM}
Since the perturbation is assumed to be small, the dynamics of
each linear mode can be studied independently. Presenting
the nonstationary solution in the vicinity of the stationary
state as follows,
\begin{equation}
\Psi({\bf r},t)=  \left[\psi_0(r) +\delta\psi({\bf r},t)\right]e^{-i\mu t}, \Phi({\bf r},t) = \Phi_0(r) + \delta \Phi({\bf r},t)
\label{eq:expand_psi_Phi}    
\end{equation}
\begin{equation}
\delta\psi({\bf r},t) = u_{l}(r)Y_{lm_z} (\theta, \varphi)e^{-i\omega t} 
+ v_{l}^*(r)Y^*_{lm_z}(\theta, \varphi)e^{i\omega^* t} 
\label{eq:expand_delta_Phi}
\end{equation}

By inserting Eqs.~(\ref{eq:expand_psi_Phi}) and (\ref{eq:expand_delta_Phi}) into Eqs.~(\ref{eq:GPE_nondim}) and (\ref{eq:Poisson_nondim}), and linearizing with respect to small perturbations, we obtain the BdG equations:
\begin{equation}
\hat{\mathcal{H}}_l u_{l}
+  \gamma N_*\psi_0^2 
v _{l}  + \psi_0 
\hat{D}*\left(\psi_0^* u_{l} + \psi_0 v_{l}\right)
= \omega u_{ l},
\end{equation}
\begin{equation}
\hat{\mathcal{H}}_l v_{l}
+  \gamma N_*\psi_0^2 
u _{l}  + \psi_0 
\hat{D}*\left(\psi_0^* u_{l} + \psi_0 v_{l}\right)
= -\omega v_{l},
\end{equation}
where $y(r)=\hat{D}*f(r)$ denotes the solutions of the radial equation:
\begin{equation}
\Delta_r^{(l)}y(r)= f(r).    
\end{equation}

We next focus on perturbations of the ground-state, spherically symmetric wave function of the solitonic core, $\psi_0(r)$. 
This wave function is real and satisfies the stationary equation (\ref{eq:StationaryGPE_dimless}).

Thus the BdG equations can be rewritten in the following form:
\begin{equation}
\label{eq:Bdg_u}
\hat{\mathcal{H}}_lu_{l}
+\hat{\chi}_l\left\{\psi_0 (u_{l}+v_{l})\right\} 
+  \gamma N_* \psi_0^2 
(u_{l}+v_{l}) = \omega u_{l},
\end{equation}
\begin{equation}
\label{eq:Bdg_v}
\hat{\mathcal{H}}_lv_{l}
+\hat{\chi}_l\left\{\psi_0 (u_{l}+v_{l})\right\} 
+  \gamma N_* \psi_0^2 
(u_{l}+v_{l}) = -\omega v_{l},
\end{equation}
where 
\begin{equation}
\hat{\chi}_l\left\{f\right\}
= -N_*\psi_0(r)\int_0^{+\infty}G_l(r,\xi)\psi_0(\xi)f(\xi)d\xi,
\end{equation}
and
\begin{equation}
G_l(r,\xi)= \frac{\xi}{2l+1}\left\{
\begin{array}{cc}
\left[{\xi}/{r}\right]^{l+1},  & 0\le \xi\le r   \\
\left[{r}/{\xi}\right]^{l} ,  & r\le  \xi < \infty
    \end{array}
    \right.
\end{equation}
Note that $\hat{\chi}_{l=0}\left\{\psi_0\right\} = \psi_0(r)\Phi_0(r)$, as it must be. While local self-interaction contributes directly to the perturbation at a given point, the gravitational response introduces a fundamentally nonlocal effect: perturbations at one location influence the potential across an extended spatial domain. This intrinsic nonlocality, encoded in the operator $\hat{\chi}_l$, necessitates a fully self-consistent integro-differential formulation of the BdG equations.

The matrix form of the BdG eigenvalue problem in dimensionless units is as follows:
\begin{equation}
   \label{eq:eqBdGmatrix}
    \left(
    \begin{array}{cc}
\mathcal{B}_{11}^{(l)}   & \mathcal{B}_{12}^{(l)}   \\
\mathcal{B}_{21}^{(l)}  &  \mathcal{B}_{22}^{(l)}
    \end{array}
    \right)
    \left(
    \begin{array}{c}
      u_{l} \\
             v_{l}
    \end{array}
    \right) = 
    \omega_{\nu l}\left(
    \begin{array}{c}
      u_{l} \\
             v_{l}
    \end{array}
    \right),
\end{equation}
where $\mathcal{B}_{11}^{(l)}=\hat{\mathcal{H}}_l + \mathcal{B}_{12}^{(l)}$, $\mathcal{B}_{12}^{(l)}=\gamma N_*\psi_0^2+ \hat{\chi}_l$, $\mathcal{B}_{21}^{(l)} =-\mathcal{B}_{12}^{(l)} $, $\mathcal{B}_{22}^{(l)} =-\mathcal{B}_{11}^{(l)}$.

The boundary conditions for eigenfunctions  for $l\ne 0$ and for $l= 0$ at the center of the solitonic core are as follows:
\begin{equation}
\label{eq:boundary_epsilon_l}
  u_{l}(0)=v_{l}(0)=0, \left.\frac{d u_{0}}{dr}\right|_{r=0}=\left.\frac{d v_{0}}{dr}\right|_{r=0}=0,
\end{equation}
and at infinity for all $l$:
\begin{equation}
\label{eq:boundary_epsilon_0}
  \lim_{r \to \infty}u_{l}(r)=\lim_{r \to \infty}v_{l}(r)=0.
\end{equation}

The dimensionless radial eigenfunctions $u_{l}(r)$ and $v_{ l}(r)$ are rescaled to satisfy the normalization condition:
\begin{equation}\label{eq:normalization_condition}
\int_0^{+\infty}  \left(\left|u_{l}(\xi)\right|^2 -\left|v_{l}(\xi)\right|^2\right) \xi^2 d\xi= \frac{1}{\psi_*^2L_*^3}=\frac{1}{N}.
\end{equation}

The BdG eigenvalue problem (\ref{eq:eqBdGmatrix}) with an integrodifferential operator can be represented in finite-difference form with zero boundary conditions $u_l(r_\mathrm{max})=0$, $v_l(r_\mathrm{max})=0$ at the endpoint of the radial grid $r=r_\mathrm{max}$. We have solved the resulting linear algebra eigenvalue problem numerically. It turns out that this approach gives an accurate description for the lower-order states for large enough $N_*$, when the eigenfunctions are well localized. However, this method is not able to accurately represent the increasingly large-scale eigenfunctions at higher $\nu$ and $l$ values since the spectrum becomes sensitive to the boundary conditions at $r_\mathrm{max}$, as was pointed out in Ref. \cite{PhysRevD.105.103506}. Thus, as the quantum numbers $\nu$ and $l$ increase in BdG equations, numerical solutions require a more sophisticated approach.

\subsection{Numerical solution of BdG problem}
In the present work, we have developed a method for solving the complete BdG problem with the integro-differential operator, which remains valid for arbitrary quantum numbers $\nu$ (number of nodes) and $l$ (orbital quantum number), including highly excited states with $\nu \gg 1$  or $l \gg 1$.

Since BdG set of equations is linear, we suggest representing the radial profiles $u_l(r)$ and $v_l(r)$ as the following linear superposition:
\begin{equation}
    \label{eq:uv_expansion}
    \left(
    \begin{array}{c}
      u_{l} (r) \\
             v_{l}(r)
    \end{array}
    \right) = \sum_{\nu = 0}^\infty\left(
    \begin{array}{c}
      U_{\nu}^{(l)}\\
             V_{\nu}^{(l)}
    \end{array}
    \right) R_{\nu l}(r),
\end{equation}
where $U_{\nu}^{(l)}$ and $V_{\nu}^{(l)}$ are constants and  functions $R_{\nu l}(r)$, form a complete orthonormal set of functions
\begin{equation}
    \label{eq:Ortohonality}
    \int_0^{+\infty}R_{\alpha l}(r)R_{\nu l}(r)r^2 dr=\delta_{\alpha \nu}.
\end{equation}

Various orthogonal bases can be used to solve the linear BdG problem. In the present work, we have used two different basis functions: (i) a basis of the hydrogen-like atom (H-like basis) and (ii) a basis of the 3D spherically symmetric quantum harmonic oscillator (QHO basis). Both solutions being converging, give the same results for the eigenstates but exhibit different speeds of convergence for different quantum numbers $\nu$ and $l$. In both cases, we choose the radial basis functions, $R_{\nu l}(r)$, in \textit{analytic form}, which allows us to accurately describe the spectrum $\omega$ and eigenfunctions $u_l(r)$, $v_l(r)$ for arbitrary quantum numbers $\nu$ and $l$. Each radial basis function 
satisfies the following ordinary differential equation:
\begin{equation}
    \label{eq:R_ODE_general}
-\frac12\Delta_r^{(l)}R_{\nu l}(r) +\Phi_\mathrm{B}(r)R_{\nu l}(r) = E_{\nu l}R_{\nu l}(r),
\end{equation}
where the trapping potential $\Phi_\mathrm{B}(r)$ for the radial basis functions is chosen to accurately describe details of the self-induced gravitational potential $\Phi_0(r)$ either in the central region  (for a basis of the 3D spherically symmetric quantum harmonic oscillator) or asymptotic behaviour for a large distance from the BEC core centre (H-like basis). The energy $E_{\nu l}$  in Eq. (\ref{eq:R_ODE_general}) does not depend on the quantum number $m_z$ due to the spherical symmetry of the potential $\Phi_\mathrm{B}(r)$. Note that the method developed here avoids issues with boundary conditions for the higher-order eigenmodes, since the vanishing at infinity basis functions automatically satisfy the appropriate asymptotic behavior required for localized eigenfunctions.

To obtain a numerical solution to the BdG problem we need to calculate the set of coefficients in expansion (\ref{eq:uv_expansion}), which form the vectors
$\mathbf{U}^{(l)} = \left\{U_{0}^{(l)}, U_{1}^{(l)}, ..., U_{\nu_\mathrm{max}}^{(l)}\right\}
$, 
$\mathbf{V}^{(l)} = \left\{V_{0}^{(l)}, V_{1}^{(l)}, ..., V_{\nu_\mathrm{max}}^{(l)}\right\}
$, 
 where $\nu_\mathrm{max}$ is chosen to guarantee a desirable accuracy in converging sum (\ref{eq:uv_expansion}). A detailed analysis of the convergence properties and optimal choice of the number of basis functions, $\nu_{\mathrm{max}}$, for both the QHO and hydrogen-like bases is provided in the Appendix.

Let us insert expansion (\ref{eq:uv_expansion}) in Eqs. (\ref{eq:Bdg_u}),(\ref{eq:Bdg_v}), accounting for Eq. (\ref{eq:R_ODE_general}) and orthogonality condition
(\ref{eq:Ortohonality}), which yields  the following linear algebraic eigenvalue problem:
\begin{equation}
\label{eq:eqBdSpectralU}
\sum_{\nu=0}^{\nu_\mathrm{max}}\left[\mathcal{H}^{(l)}_{\alpha\nu} U_{\nu}^{(l)}+
\mathcal{K}^{(l)}_{\alpha\nu}\left(U_{\nu}^{(l)}+V_{\nu}^{(l)}\right)\right]=\omega U_{\alpha}^{(l)}
\end{equation}
\begin{equation}
\label{eq:eqBdSpectralV}
\sum_{\nu=0}^{\nu_\mathrm{max}}\left[\mathcal{H}^{(l)}_{\alpha\nu} V_{\nu}^{(l)}+
\mathcal{K}^{(l)}_{\alpha\nu}\left(U_{\nu}^{(l)}+V_{\nu}^{(l)}\right)\right]=-\omega V_{\alpha}^{(l)}
\end{equation}
where $\mathcal{H}^{(l)}_{\alpha\nu}$ is the matix element of operator $\hat{\mathcal{H}}^{(l)}$ defined in Eq. (\ref{eq:H_l_equation}):
\begin{eqnarray}
\nonumber
\label{eq:defHmatrix}
\mathcal{H}^{(l)}_{\alpha\nu}=\int_0^{+\infty}r^2R_{\alpha l}(r)\hat{\mathcal{H}}_lR_{\nu l}(r) dr
\\ =(-\mu +E_{\alpha l})\delta_{\alpha \nu}+\mathcal{F}_{\alpha \nu}^{(l)}+ \mathcal{P}_{\alpha \nu}^{(l)}, 
\end{eqnarray}
where $\delta_{\alpha \nu}$ is the Kronecker delta and 
\begin{equation}
\label{eq:F}
\mathcal{F}_{\alpha \nu}^{(l)}= \int_0^{+\infty}\left[\Phi_0(r)-\Phi_\mathrm{B}(r)\right]R_{\alpha l}(r)R_{\nu l}(r)r^2 dr,
\end{equation}
\begin{equation}
\label{eq:P}
\mathcal{P}_{\alpha \nu}^{(l)}=\gamma N_*\int_0^{+\infty}\psi_0^2(r)R_{\alpha l}(r)R_{\nu l}(r)r^2 dr.
\end{equation}
The matrix $\hat{\mathcal{K}}^{(l)}$ in Eqs. (\ref{eq:eqBdSpectralU}), (\ref{eq:eqBdSpectralV}) is defined as follows:
\begin{equation}
\label{eq:defK}
\mathcal{K}^{(l)}_{\alpha\nu}= \mathcal{P}_{\alpha \nu}^{(l)}- \mathcal{Q}_{\alpha \nu}^{(l)},
\end{equation}
where
\begin{equation}
\label{eq:Q_alpha_nu}
\mathcal{Q}_{\alpha \nu}^{(l)}= N_*\int_0^{+\infty}\int_0^{+\infty}q_{\alpha\nu}^{(l)}(r,\xi)r^2 dr d\xi,
\end{equation}
\begin{equation}
    q_{\alpha\nu}^{(l)}(r,\xi)=\psi_0(r)\psi_0(\xi)G_l(r,\xi)R_{\alpha l}(r)R_{\nu l}(\xi).
\end{equation}

Let us rewrite BdG equations using the symmetric and antisymmetric superposition of the eigenvectors: $\mathbf{S}=\mathbf{U}^{(l)}+\mathbf{V}^{(l)}$ and $\mathbf{A}=\mathbf{U}^{(l)}-\mathbf{V}^{(l)}$ as follows:
$\hat{B}_- \mathbf{A}=\omega \mathbf{S}$ and $\hat{B}_+ \mathbf{S}=\omega \mathbf{A}$, where $(\hat{B}_-)_{\alpha \nu}={\mathcal{H}}^{(l)}_{\alpha \nu}$ and 
$(\hat{B}_+)_{\alpha \nu}={\mathcal{H}}^{(l)}_{\alpha \nu} + 2{\mathcal{K}}^{(l)}_{\alpha \nu}$.
Finally we obtain two decoupled linear eigensystems for  $\mathbf{A}$ and $\mathbf{S}$:
\begin{eqnarray}
\label{eq:decoupled_BdG}
\hat{B}_+\hat{B}_- \mathbf{A}= \omega^2 \mathbf{A}, \,\, 
 \hat{B}_-\hat{B}_+\mathbf{S}= \omega^2 \mathbf{S}.
\end{eqnarray}
These eigensystems have been solved numerically for various values of the parameter $N_*$.


\begin{figure}[htb]
    \centering
    \includegraphics[width=0.75\textwidth]{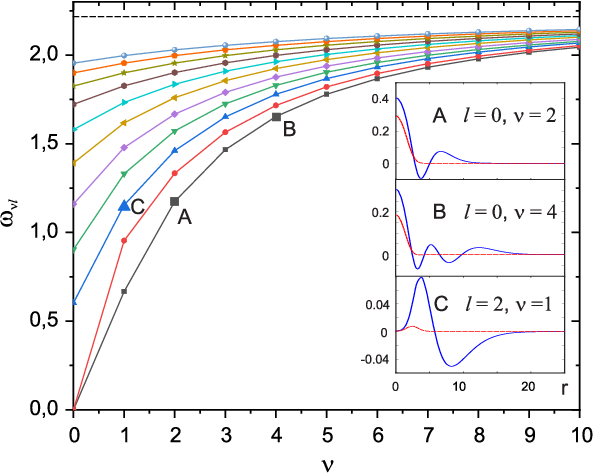}
   \caption{Eigenfrequencies $\omega_{\nu l}$ for $l\in [0,10]$, calculated numerically from the Bogoliubov-de Gennes equations for $N_*=100$.  The inset shows examples of the eigenfunctions (in arbitrary units) for different values of quantum numbers $\nu$ and $l$.}
    \label{fig:uvL}
\end{figure}

\subsection{Excitation spectrum}
Numerical results for the eigenvalues are presented in Fig. \ref{fig:uvL} for 
$N_*=100$. The inset of Fig.\ref{fig:uvL} displays typical eigenfunction profiles 
$u_l(r)$ and $v_l(r)$ for 
$l=0$ and $l=2$ and illustrating different numbers of nodes in the radial profiles. The corresponding eigenvalues are denoted by capital-letter symbols in Fig.~\ref{fig:uvL}. Arbitrary units are used for the eigenfunctions to avoid the very small values imposed by the normalization condition (\ref{eq:normalization_condition}). 

All eigenfrequencies of the solitonic ground-state perturbations are real, confirming the well-established stability of ground-state solitonic configurations under repulsive self-interaction (see, e.g., \cite{chavanis_book,PhysRevD.109.104011}). 
The eigenfrequencies of highly excited states with $\nu \gg 1$ and $l \gg 1$ are well approximated by $\omega_{\nu l} \approx -\mu + E_{\nu l}$, where $E_{\nu l}$ represents the energy of a hydrogen-like system:
\begin{equation}
\label{eq:E_nu_l_Hlike}
E_{\nu l} = -\frac{Z^2}{2(\nu + l + 1)^2}.
\end{equation}
In the limit $\nu \gg 1$ and $l \gg 1$, $\omega_{\nu l} \to -\mu$, as illustrated in Fig. \ref{fig:uvL}. The black dashed line in the figure indicates the asymptotic value $-\mu$.

The lowest-energy solution is given by the wave functions $u_{0 0}(r)=\psi_0(r)$, $v_{0 0}(r)=-\psi_0(r)$, with the eigenvalue $\omega_{0 0}=0$. Notably, the dipole-like $l=1$ node-less mode $\nu=0$ also corresponds to the zeroth frequency. However, this mode does not represent an excitation of the ground state but rather produces a center-of-mass motion, and we do not consider this type of perturbation in the present work.

The lowest non-zero-frequency mode corresponds to the eigenstate with $l=0$ and a single node ($\nu=1$). This mode represents periodic breathing, characterized by low-frequency, radially symmetric oscillations of the solitonic core. We further investigate these oscillations in detail and compare our numerical results with predictions from a simple variational approach.

\textcolor{black}{Our BdG  analysis provides a unified and self-consistent framework for modeling the linear response of solitonic cores in ultralight bosonic dark matter (ULDM) halos. Each eigenmode of the BdG equations corresponds to a specific collective excitation channel, representing a possible oscillatory mode that the self-gravitating condensate can exhibit following a small perturbation. This includes breathing, quadrupole, and higher-order deformations of both the density and the gravitational potential. In contrast to time-dependent simulations, which capture nonlinear dynamics but often obscure individual mode contributions, the BdG approach resolves the excitation spectrum into well-defined eigenfrequencies and mode shapes. This provides a physically transparent interpretation of ULDM core dynamics and enables systematic identification of astrophysically relevant effects such as dynamical heating and time-dependent potential fluctuations.}


In addition to capturing collective excitation spectra, it enables a quantitative study of the thermodynamic properties of self-gravitating ULDM condensates, including both quantum depletion at zero temperature and thermal effects at finite temperature. Condensate depletion, a well-established feature of atomic Bose–Einstein condensates with contact or long-range dipole–dipole interactions~\cite{PhysRevA.86.063609}, can be quantified in the gravitational context using our BdG results. For realistic halo parameters, the gas parameter is extremely small ($n a_s^3 \ll 1$), indicating that the noncondensed fraction at $T=0$ (quantum depletion) remains negligible, consistent with atomic BEC predictions~\cite{PhysRevA.88.053617}.

At finite temperature, the BdG excitation spectrum enables accurate computation of the thermal component and the critical temperature for condensation in localized, self-gravitating halos~\cite{PhysRevLett.78.1842, giorgini1997thermodynamics}. In particular, this approach provides a physically grounded route to modeling the transition between the solitonic core and the surrounding isothermal halo, which is crucial for realistic finite-temperature descriptions of galactic dark matter structures. These insights lay the groundwork for future studies of how thermal excitations and core oscillations may perturb stellar systems or influence the long-term stability of galactic cores.

\section{Breathing excitations of the solitonic core}
In this section, we study the compressible mode $\nu=1$, $l=0$, which represents low-frequency oscillations of the solitonic core. 
To provide physical context and enable quantitative comparison, we present known analytical expressions for the breathing-mode frequency, obtained using well-established variational approaches: the Hamiltonian method with a Gaussian ansatz~\cite{PhysRevD.84.043531} and the Lagrangian formalism~\cite{PhysRevD.94.083007}. These analytical results serve as useful reference points for validating our full numerical BdG analysis.

\begin{figure}[htb]
    \centering
    \includegraphics[width=\textwidth]{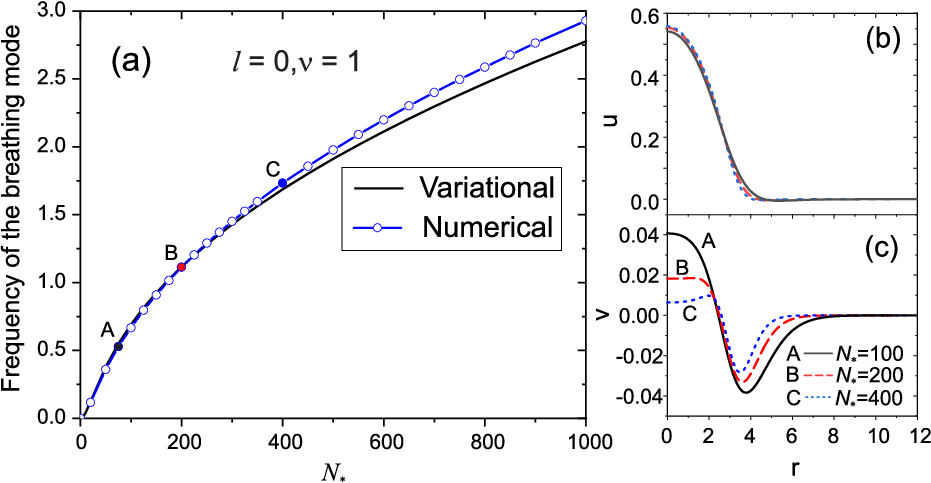}
   \caption{(a) Breathing compressible mode $\nu=1$, $l=0$  frequency, $\omega_{1, 0}$, vs $N_*$ 
 obtained numerically (blue line with circles) and in variational approximation (solid black line).  Eigenfuncions (b) $u(r)$ and (c) $v(r)$ for three different values of the parameter $N_*$.}
    \label{fig:breathing_mode}
\end{figure}

Here we use a standard variational approach (see, e.g., \cite{malomed2022soliton,chavanis_book}) with a normalized Gaussian trial function in dimensionless form: 
\begin{equation}
\Psi(\mathbf{r},t) = \frac{1}{\sqrt{\pi^{3/2}R^3(t)}} e^{-\frac{r^2}{2R^2(t)} + i\beta(t)r^2},
\label{eq:variat_trialfunction}
\end{equation}
where $R(t)$ and $\beta(t)$ are time-dependent variational parameters.
The corresponding Lagrangian is given by
\begin{equation}
\mathcal{L} = -\frac{3}{2} \left[ R^2 \Dot{\beta} + \frac{1}{2R^2} + 2\beta^2R^2 \right] + \frac{N_*}{4\sqrt{2}\pi^{3/2}R^3} \left( 1 - R^2 \right).
\label{eq:lagrangian}
\end{equation}
Differentiating Lagrangian (\ref{eq:lagrangian}) one obtains the first Euler-Lagrange equation $\frac{\partial \mathcal{L}}{\partial R} - \frac{d}{dt}\frac{\partial \mathcal{L}}{\partial \dot{R}} =0$:
\begin{equation}\label{eq:FirstLagrange}
  -3R\dot{\beta} + \frac{3}{2 R^{3}}  - 6\beta^{2}R  - \frac{N_{*}}{8\pi R^{2}}\sqrt{\frac{2}{\pi}}  + \frac{3N_{*}}{2(2\pi)^{3/2}R^{4}} = 0.
\end{equation}

The second Euler-Lagrange equation reads $\frac{\partial \mathcal{L}}{\partial \beta} - \frac{d}{dt}\frac{\partial \mathcal{L}}{\partial \dot{\beta}} =0 $:
\begin{equation}\label{eq:beta}
  \beta  = \frac{\dot{R}}{2R}.
\end{equation}

Substituting Eq. (\ref{eq:beta}) into Eq. (\ref{eq:FirstLagrange}), we obtain:
\begin{equation}
  -\frac{3\ddot{R}}{2}  + \frac{3}{2 R^{3}}  -  \frac{N_{*}}{8\pi R^{2}}\sqrt{\frac{2}{\pi}}  + \frac{3N_{*}}{2(2\pi)^{3/2}R^{4}} = 0.
\end{equation}

The solution of the last equation in time-independent case is ground state with $\beta =  0$ and $R = R_{0}$. In the vicinity of stationary solution variational parameter has a form  $R(t) = R_{0} + \delta(t)$. In the first order, we obtain:
\begin{equation*}
  \ddot{\delta} + \omega^2\delta = 0,
\end{equation*}
where 
\begin{equation}
    \omega^{2} = \frac{3}{R_{0}^{4}}  - \frac{N_{0}}{6\pi R_{0}^{3}}\sqrt{\frac{2}{\pi}}  + \frac{4N_{0}}{(2\pi)^{3/2}R_{0}^{5}} = \frac{2}{3N_{0}}\frac{\partial^{2}E}{\partial R^{2}}\vert_{R = R_{0}},
\end{equation} the energy functional is as follows:
\begin{equation}
E=\frac{3}{4R^2}+\frac{N_*}{4\sqrt{2}\pi^{3/2}R^3}\left(1-R^2\right).
    \label{eq:energy_variat}
\end{equation}
The stationary state corresponds to the minimum of the energy: $\frac{\partial E}{\partial R}\left.\right|_{R=R_0}=0$, where 
\begin{equation}
R_0=\frac{3\sqrt{2\pi^3}}{N_*}\left[1+\sqrt{1+\frac{N_*^2}{6\pi^3}}\right].    
\end{equation}
 The breathing mode frequency is given as follows:
\begin{equation}
    \omega_b^2=\frac{2}{3N_*}\left.\frac{\partial^2E}{\partial R^2}\right|_{R=R_0}.
\end{equation}



Figure~\ref{fig:breathing_mode}(a) presents the breathing-mode frequency as a function of the normalized solitonic mass. The variational approach (solid black curve) agrees well with the numerical solution of the BdG equations (blue circles), with deviations appearing only at large $N_*$, where the system enters the Thomas-Fermi (TF) regime and the local self-interaction dominates over the kinetic term. In this regime, the Gaussian variational ansatz [Eq.~(\ref{eq:variat_trialfunction})] becomes less accurate in capturing the radial structure of the BEC core.

Figures~\ref{fig:breathing_mode}(b) and \ref{fig:breathing_mode}(c) show the eigenfunctions $u(r)$ and $v(r)$ of the breathing mode. For small $N_*$, the profile of $v(r)$, which has one node, exhibits a central maximum. As $N_*$ increases, the profile becomes distorted and develops a local minimum near the center, reflecting the increasing role of repulsive self-interaction as the system transitions to the TF regime.

\textcolor{black}{
Let us now compare the breathing-mode frequency obtained from our BdG analysis with results of cosmological simulations of ULDM halos. In our approach, the breathing mode appears as the lowest spherically symmetric ($l=0$, $\nu=1$) excitation around the stationary soliton solution of the GPP system, calculated from the linearized BdG spectrum. In contrast, cosmological simulations~\cite{PhysRevD.98.043509,PhysRevD.103.103019,PhysRevD.108.063519} observe oscillatory behavior similar to that of an emerging nonlinear phenomenon: solitonic cores develop large-scale pulsation following halo mergers, tidal perturbations, or soliton formation events. These studies track the evolution of the core density or gravitational potential over time and extract the dominant frequency either by direct time-series analysis or via Fourier decomposition of density fluctuations. Despite the methodological differences, the frequency range is consistent. For example, a soliton with $N_*=400$ and particle mass $m = 10^{-22} \,\mathrm{eV}$ yields a breathing frequency $\tilde{\omega}_b \approx 10^{-14} \,\mathrm{s}^{-1}$ in our calculation, corresponding to a timescale of several hundred million years, similar to the oscillation periods recovered in~\cite{PhysRevD.103.103019,PhysRevD.108.063519}. This agreement supports the interpretation of the BdG breathing mode as the coherent pulsation seen in simulations and highlights its relevance for gravitational relaxation and energy redistribution in evolving ULDM halos. In the following section, we discuss the broader astrophysical implications of these oscillations, including their potential role in core-halo interactions and observable signatures.
}

\section{Astrophysical Implications of Solitonic Core Oscillations}
Collective excitations of solitonic cores can be triggered by a range of realistic astrophysical processes and may significantly influence the dynamical evolution of ULDM halos. Strong perturbations arise during halo mergers, where central solitons are distorted and internal modes are excited. Tidal interactions, baryonic infall, and dynamical friction similarly generate time-dependent gravitational fields that couple to the core. Even in isolated systems, stochastic fluctuations from early structure formation or an incoherent ULDM background can drive low-amplitude oscillations over cosmological timescales. When these perturbations remain small compared to the binding energy of the core, the internal response can be treated within linear theory. The collective modes analyzed in this work define the intrinsic response channels of the soliton and provide a self-consistent framework to interpret its dynamics under realistic conditions, with potential signatures in simulations and observations of astrophysical systems.

Once excited, these internal oscillations can induce fluctuations in the gravitational potential that may lead to dynamical heating of embedded star clusters in dwarf galaxies, altering their velocity dispersions over gigayear timescales~\cite{PhysRevD.103.023508}. This mechanism has been discussed in the context of Eridanus II, where constraints on soliton-induced heating offer valuable bounds on the mass of ULDM particles~\cite{PhysRevD.103.103019}. These effects are qualitatively expected from the stochastic gravitational potential perturbations generated by the granular density distribution of fuzzy dark matter~\cite{PhysRevD.95.043541}. Similar dynamical relaxation processes have been seen in the scalar field collapse simulations, where solitonic cores shed energy and approach equilibrium via gravitational cooling~\cite{PhysRevD.69.124033}. 
Moreover, numerical simulations of core-tail systems have shown that gravitational interaction between a solitonic core and its surrounding envelope can excite long-lived breathing-mode oscillations and drive the system toward virialized states~\cite{PhysRevD.108.063519}.
In addition to dynamical effects on embedded stellar systems, oscillating solitonic cores may induce low-amplitude, time-dependent perturbations in the gravitational field. Although the amplitude of these gravitational potential oscillations is small, they may, in principle, produce weak lensing distortions. Such wave-induced potential fluctuations are characteristic of the quantum pressure terms appearing in the Schr\"odinger–Poisson formulation of ULDM~\cite{PhysRevD.95.043541}. However, these lensing effects are expected to lie well below current observational sensitivities and would likely require future missions with ultra-high precision.
On larger scales, the interaction between solitonic cores and halo environments may affect the evolution of halo density profiles and the survival of substructures. Cosmological simulations have shown that repulsive self-interactions can influence soliton formation, stability, and their coupling to the surrounding halo~\cite{PhysRevD.109.043516}, and that core–halo interactions can lead to long-term evolution of solitonic cores and distortions in rotation curves~\cite{PhysRevD.110.063502}. These effects may lead to modulations in the inner halo structure and influence the merger history of subhalos. Furthermore, breathing modes of the solitonic core could induce deviations in galactic rotation curves, potentially distinguishable from standard cold dark matter predictions. Such effects are also supported by predictions for soliton-halo relaxation times and core-envelope transitions~\cite{PhysRevD.95.043541}.
The presence of a supermassive black hole at the center of a solitonic core can strongly deform its density profile, especially within the inner parsecs. The resulting anisotropic potential affects stellar orbits and compresses the core, inducing transient deviations from virial equilibrium~\cite{PhysRevD.100.083022}. Semi-analytical and numerical studies further show that compact central objects, including black hole, can significantly reshape solitons~\cite{PhysRevD.111.023006}, offering a potential probe via stellar kinematics. While the present work does not model these effects explicitly, the computed excitation spectrum provides a theoretical basis for analyzing such processes in future simulations and semi-analytic frameworks.

Taken together, these dynamical phenomena associated with galactic cores emphasize the importance of both future observations and theoretical modeling targeting the inner regions of galaxies to constrain the properties of ultralight bosons and test the viability of ULDM as a dark matter candidate.

\section{Summary and conclusions}

We have demonstrated that ultralight bosonic dark matter halos host solitonic cores exhibiting rich dynamical behavior characterized by collective oscillation modes. These modes can leave distinct observational signatures in galactic dynamics. 

To characterize these excitations, we implemented a fully self-consistent numerical solution of the Bogoliubov–de Gennes equations, accounting for both gravitational potential fluctuations and local self-interactions. The formalism, which describes linear perturbations around a stationary solitonic background, accurately captures the spectrum of small collective excitations. Within the framework developed in this work, we solve the full non-local integro-differential eigenvalue problem that governs the system, allowing a detailed analysis of the mode structures and their impact on the density profile and gravitational field of the ULDM core. We provide example parameter mappings to physical units, which demonstrate consistency with prior theoretical and numerical studies of ultralight dark matter solitons. The computed excitation spectrum offers a quantitative framework to constrain ULDM particle properties through dynamical heating of embedded stellar systems and time-dependent density and potential fluctuations in halo simulations.

Moreover, our results establish a solid basis for studying thermodynamic properties in gravitationally localized Bose–Einstein condensates. This approach allows for accurate exploration of finite-temperature effects, including the determination of critical temperatures and quantitative analysis of the isothermal nondegenerate halo. Such thermodynamic investigations are essential for understanding realistic physical scenarios of solitonic formation and evolution and provide important insights into the physical conditions within galactic halos formed by ultralight dark matter, stimulating future observational and theoretical studies.


\section*{Acknowledgements}
The authors thank E. Gorbar and K. Korshynska for useful discussions.
The authors are partially supported by “Iniziativa Specifica Quantum” of INFN and by the Project "Frontiere Quantistiche" (Dipartimenti di Eccellenza) of the Italian Ministry for Universities and Research. LS is partially supported by funds of the European Union - Next Generation EU: European Quantum Flagship Project "PASQuanS2", National Center for HPC, Big Data and Quantum Computing [Spoke 10: Quantum Computing].  AY is supported by the PRIN Project "Quantum Atomic Mixtures: Droplets, Topological Structures, and Vortices".

\section{Appendix}

 \subsection{A basis of the hydrogen-like atom}
As was pointed out the wave function $\psi_0(r)$ of the fundamental soliton is localized and its profile is well-described in TF approximation for large values of $N_*$ so that for $r\gg r_{TF}$ $\psi_0\to 0$ and the gravitation potential tends to Coulomb asymptotic so that trapping potential in Eq. (\ref{eq:R_ODE_general}) is defined as follows: $\Phi_\mathrm{B}(r) = -Z/r$ with $Z=N_*/(4\pi)$.  
Thus the natural choice of basis functions is the set of the radial wave functions of the hydrogen-like atom (H-like basis):
\begin{equation}
    \label{eq:Rbasis}
    R_{\nu l}(r) = C_{\nu l} e^{-\rho/2}\rho^l L_\nu^{(2l+1)}(\rho),
\end{equation}
and normalization constant, $C_{\nu l}$,  is found from the normalization condition (\ref{eq:Ortohonality}):
\begin{equation}
    C_{\nu l} = \sqrt{\frac{4Z^3\nu!}{(\nu+l+1)^4(\nu+2l+1)!}}
\end{equation}
where $\rho=\frac{2Zr}{\nu+l+1}$. Here $L_\nu^{(2l+1)}(\rho)$  is a generalized Laguerre polynomial of degree $\nu$, which defines the number of nodes of the radial profile. 

We adopt the hydrogen-like atom basis, constructed from the eigenfunctions of the linear Schr\"odinger equation with a Coulomb potential of effective nuclear charge \( Z \), to model the gravitational potential in the solitonic core. This choice captures the correct asymptotic behavior of the potential at large distances, where the soliton’s self-gravity approaches a \( 1/r \) form.

The radial basis functions $R_{\nu l}(r)$ satisfy the following differential equation:
\begin{equation}
    \label{eq:R_ODE}
    -\frac{1}{2}\Delta_r^{(l)} R_{\nu l}(r) - \frac{Z}{r} R_{\nu l}(r) = E_{\nu l} R_{\nu l}(r),
\end{equation}
where the energy levels $E_{\nu l}$ depend on the principal quantum number $n = \nu + l + 1$, as given in Eq.~(\ref{eq:E_nu_l_Hlike}).


The hydrogen-like atom basis gives very accurate description of the higher-order states of the BdG system for $\nu\gg 1$ and $l\gg 1$. However, the singular behaviour of the Coulomb potential leads to slow convergence of the expansion for eigenfunctions with $\nu=0$ or $l=0, 1$.

\subsection{Harmonic oscillator basis}
The gravitational potential for the lowest-order states can be well approximated by the parabolic potential. Indeed, in TF approximation for $r\ll R_{\mathrm{TF}}$ the potential $\Phi_{\mathrm{TF}}(r)=-\frac{N_*}{4\pi^2}(1+\frac{1}{r}\sin r)\approx \Phi_{\mathrm{TF}}(0) +\frac{1}{2}\Omega^2 r^2$, where $\Phi_{\mathrm{TF}}(0)= -\frac{N_*}{2\pi^2}$, $\Omega^2=N_*/(12\pi^2)$. Thus, the trapping potential in Eq. (\ref{eq:R_ODE_general}) can be taken as follows: $\Phi_\mathrm{B}(r) =\frac{1}{2}\Omega^2 r^2$. The basis of 3D spherically-symmetric  oscillator functions:
\begin{equation}
    \label{eq:Rbasis_oscil}
    R_{\nu l}(r) = C_{\nu l} e^{-\rho^2/2}\rho^l L_\nu^{(l+1/2)}(\rho^2),
\end{equation}
\begin{equation}
    C_{\nu l}=2^{\nu+l+1}\left(\Omega^3/\pi\right)^{1/4}\sqrt{\frac{\nu!(\nu+l)!}{(2\nu+2l+1)!}}
\end{equation}
where $\rho=\sqrt{\Omega}r,$  normalization constant is found from the normalization condition (\ref{eq:Ortohonality}). Here $L_\nu^{(2l+1)}(\rho)$  is a generalized Laguerre polynomial of degree $\nu$, which defines the number of nodes of the radial profile. 

The corresponding potential $\Phi_\mathrm{B}(r)=\frac{1}{2}\Omega^2r^2$ and the energy $E_{\nu l}$ in Eqs. (\ref{eq:R_ODE_general}), (\ref{eq:defHmatrix}), and (\ref{eq:F}) is as follows:
\begin{equation}
    \label{eq:E_nu_l}
    E_{\nu l}=\Omega\left(2\nu +l + \frac{3}{2}\right).
\end{equation}

\begin{figure}[htb]
    \centering
    \includegraphics[width=0.75\textwidth]{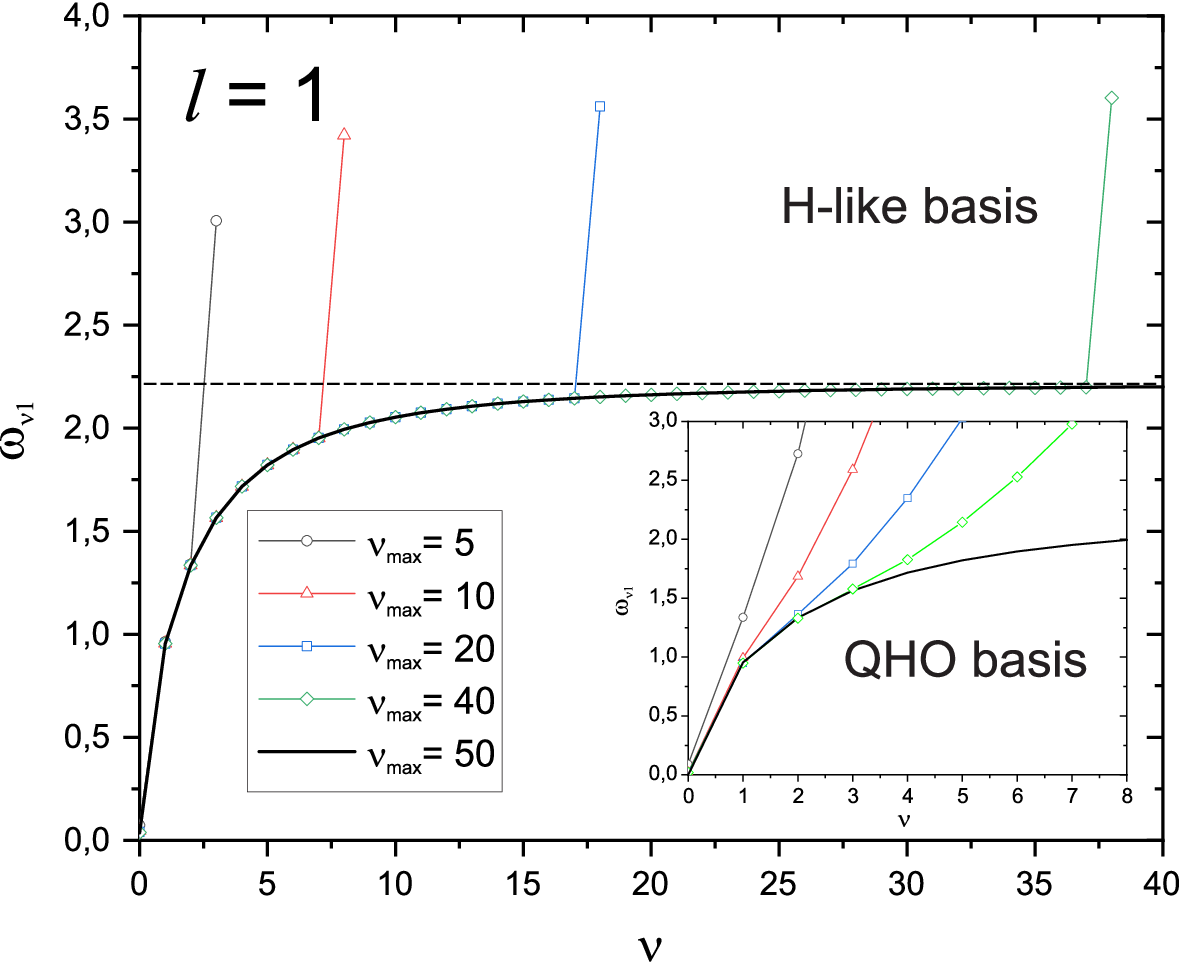}
   \caption{Illustration of convergence of the eigenvalues $\omega_{\nu l=1}$ for different numbers of the basis functions $\nu_\mathrm{max}$ using a hydrogen-like atom (H-like) basis. The inset demonstrates the convergence using a quantum harmonic oscillator (QHO) basis, highlighting the faster convergence rate of the hydrogen atom basis.}
    \label{fig:convergence}
\end{figure}
\subsection{Basis comparison and convergence analysis}
As discussed in the main text, obtaining a numerical solution to the BdG problem requires computing the coefficients in the expansion (\ref{eq:uv_expansion}), where \(\nu_\mathrm{max}\) is chosen to ensure the desired accuracy in the convergence of the sum (\ref{eq:uv_expansion}). We have tested the convergence properties and the optimal selection of the number of basis functions for both the harmonic oscillator and hydrogen-like atom bases.

Figure \ref{fig:convergence} presents our findings for the mode $l=1$. The solid curve shows the numerically determined eigenfrequencies of the modes with \(\nu\) nodes in the radial profile of $\nu_\mathrm{max} = 50$ in (\ref{eq:uv_expansion}), using the basis of hydrogen-like atoms. The colored open symbols correspond to the eigenfrequencies obtained with restricted bases for various values of $\nu_\mathrm{max}$. The inset displays the same results for the harmonic oscillator basis. 

A similar analysis was carried out for different values of the quantum number $l$, which gives a similar result for convergence. As shown in Fig. \ref{fig:convergence}, both bases converge to the same eigenvalues $\omega_{\nu l}$. However, the hydrogen-like atom basis exhibits significantly faster convergence, requiring fewer terms in the expansion (\ref{eq:uv_expansion}) to accurately reproduce the spectrum.

\bibliography{Refs}

\end{document}